\documentclass[aps,footinbib,twocolumn,showpacs,amsmath,amssymb,superscriptaddress,prb]{revtex4}
\usepackage{graphicx,amsmath}
\usepackage{epstopdf}
\usepackage{amssymb}
\usepackage{grffile}
\usepackage[usenames]{color}
\usepackage{indentfirst}
\usepackage{float}
\usepackage{color}
\usepackage{mathrsfs}
\usepackage{dcolumn}
\usepackage{bm}
\usepackage[colorlinks=true,bookmarks=false,citecolor=blue,linkcolor=red,urlcolor=blue]{hyperref}

\begin{document}
\title{Dynamics of fractional quantum Hall Liquids with a pulse at the edge}
\author{Jie Li}
\affiliation{Department of Physics and Chongqing Key Laboratory for Strongly Coupled Physics, Chongqing University, Chongqing 401331, People's Republic of China}
\author{Chen-Xin Jiang}
\affiliation{Department of Physics and Chongqing Key Laboratory for Strongly Coupled Physics, Chongqing University, Chongqing 401331, People's Republic of China}
\affiliation{Division of Physics and Applied Physics, Nanyang Technological University, Singapore 637371}
\author{Zi-Xiang Hu}
\email{zxhu@cqu.edu.cn}
\affiliation{Department of Physics and Chongqing Key Laboratory for Strongly Coupled Physics, Chongqing University, Chongqing 401331, People's Republic of China}

\begin{abstract}
Motivated by recent experimental advancements in scanning optical stroboscopic confocal microscopy and spectroscopy measurements, which have facilitated exceptional energy-space-time resolution for investigating edge and bulk dynamics in fractional quantum Hall systems, we formulated a model for the pump-probe process on the edge. Starting with a ground state, we applied a tip potential near the fractional quantum Hall liquid edge, which was subsequently turned off after a defined time duration. By examining how the specific nature of the tip potential influences the evolution of the wave function and its distribution in energy spectrum, we identify that quench dynamics of the edge pulse leads to excitations that spread both along the edge and perpendicularly into the bulk. Moreover, magnetoroton excitations are predominant among the bulk excitations. These results align well with the experimental observations.  Furthermore, we analyzed the effects of the tip's position, intensity, and duration on the dynamics.
 \end{abstract}
\date{\today}
\maketitle

\section{Introduction}
The fractional quantum Hall effect (FQH), distinguished by topologically protected gapped bulk states and gapless chiral edge states, has become an important framework for investigating novel quantum phenomena within strongly correlated topological systems~\cite{Tsui, Laughlin83}. This contrasts with its integer counterpart, which features as noninteracting state. Due to the exotic properties of the FQH phase, such as fractional charge, fractional statistics, and topological order, its potential applications in areas such as topological quantum computing have attracted widespread attention~\cite{Wilczek,Halperin,Arovas,Kitaev,NayakRMP}. The FQH state is characterized by compressible metallic edge states and a bulk that is incompressible and insulating. Since the magnetic field breaks time-reversal symmetry, edge modes propagate in only one direction, and conduction is limited to the edges, which can theoretically be described by chiral Luttinger liquid~\cite{Wen92,Wen90}. In contrast to the edges, the bulk of the FQH liquid is topologically protected which is immune to any local perturbation. Thus, the excitation in the bulk is also gapped. Girvin \textit{et al.} introduced the single-mode approximation (SMA) to describe the low-energy excitations in the bulk, which are neutral magneto-roton excitations~\cite{Girvin85,Girvin86,YangBo2021}.Detecting neutral excitation experimentally poses significant challenges. The bulk-edge correspondence implies that edge modes are typically vital for examining the topological properties of FQH liquids.~\cite{Chandran,Z.X.Luo,Manfra2018,Shtrikman,NakamuraNP19,NakamuraNP20,Feve, MacDonald1990,Wen1994,Wan2003,Rezayi2002,Hu2011,Mahalu2017, Kane1992,Wen1993,Kane1994,Moon1993,XLin2012, Mahalu2010}.

Excitations at the edge are also known as edge magnetoplasmons (EMPs) or charge density waves, where voltage pulses applied at the edge of a Hall liquid are converted into EMP wave packets and transmitted along the edge adjacent to the injection pulse~\cite{West,Harris,Eberl,Glazman,Klitzing}. In general, edge excitations dominate the system's low-energy behavior and possess rich physics, such as anyon statistics~\cite{Wilczek,Manfra2018,Shtrikman,NakamuraNP19,NakamuraNP20,Feve}, edge reconstruction~\cite{MacDonald1990,Wen1994,Wan2003,Rezayi2002,Hu2011,Mahalu2017}, edge tunneling~\cite{Kane1992,Wen1993,Kane1994,Moon1993,XLin2012}, and charge-neutral upstream Majorana modes of edge currents~\cite{Mahalu2010}. Due to chiral edge modes with edge velocity~\cite{Hu09}, the properties of edge states are typically measured using shot noise~\cite{Mahalu2017,Mahalu2010,Umansky2019}, and thermal transport~\cite{Yacoby,Stern,Umanskyheat,Berg} which are basically static measurements.  Recent experimental advances in scanning optical stroboscopic confocal microscopy and spectroscopy ~\cite{Yusa2013, Yusa2022, Yusa2023,Yusa2025} have enabled unprecedented energy-space-time resolution to probe edge and bulk dynamics in FQH systems. For example, pump probe reflectance measurements with $\sim 1$ps temporal resolution revealed distinct EMP modes and nonlinear excitations propagating at velocities of $\sim 10^4 - 10^5$m/s at $\nu = 1/3$, while time-resolved photoluminescence (PL) spectroscopy highlighted the role of trion lifetimes in limiting temporal resolution (100–300 ps) for edge-state imaging. These studies demonstrated that voltage pulses applied to gate electrodes can generate chiral EMPs whose propagation dynamics reflect the Tomonaga-Luttinger liquid behavior of edge channels. Complementary experiments using spatially resolved PL microscopy visualized bulk magnetoroton excitations ($10^3-10^4$ m/s) and strain pulses, highlighting the interaction between edge and bulk collective modes in the $\nu = 2/3$ FQH regime~\cite{Yusa2023,Yusa2025}. Notably, perturbations near the edge, such as gate-induced charge density modulations, were shown to excite both chiral edge waves and bulk modes, with the latter exhibiting velocity dependencies tied to the dielectric environment and Landau-level mixing. However, the microscopic mechanisms that govern the response of the edge to localized potentials, such as tip-induced confinement or disorder, remain poorly understood. Numerical studies of edge dynamics under tailored tip potentials could bridge this gap, offering insights into edge reconstruction, nonlinear excitations, and emergent spacetime metrics predicted in quantum gravity analogs. By validating experimental observations and extending them to atomistic or field-theoretic models, such simulations may elucidate how localized perturbations modify edge-state coherence, fractional statistics, and energy transport, which are critical for advancing FQH-based quantum technologies and fundamental physics.

In this work, we numerically investigate the quench dynamics of a $\nu = 1/3$ Laughlin state under a time-limited tip potential at the edge. By solving the time-dependent Schrödinger equation for a disk geometry with tunable pulse parameters (position $w$, duration $\tau$, and strength $U_{\delta}$), we demonstrate that localized perturbations induce (i) chiral edge currents, (ii) bulk diffusion mediated by magnetoroton excitations, and (iii) oscillations in quantum fidelity governed by energy gaps in the rotor spectrum. Our simulations reveal that pulse positioning near electron density maxima enhances bulk-state hybridization, while pulse duration and strength modulate excitation amplitudes through interference effects. These results establish a framework for engineering edge-bulk coupling in FQH systems and provide insights into nonlinear dynamics predicted for quantum spacetime analogs. The remainder of this paper is organized as follows. In Section II, we introduce the microscopic model used in this work and describe how to simulate voltage pulses with a tip potential. In Section III, we discuss the quench dynamics of the edge within the model Hamiltonian. In Section IV, we analyze in detail the factors that affect edge dynamics, including the position, duration, and intensity of the pulse at the edge. Summaries and discussion are given in Section V.

\section{Microscopic Model} \label{sec:model}
\begin{figure}[ht] 
  \includegraphics[width=8cm]{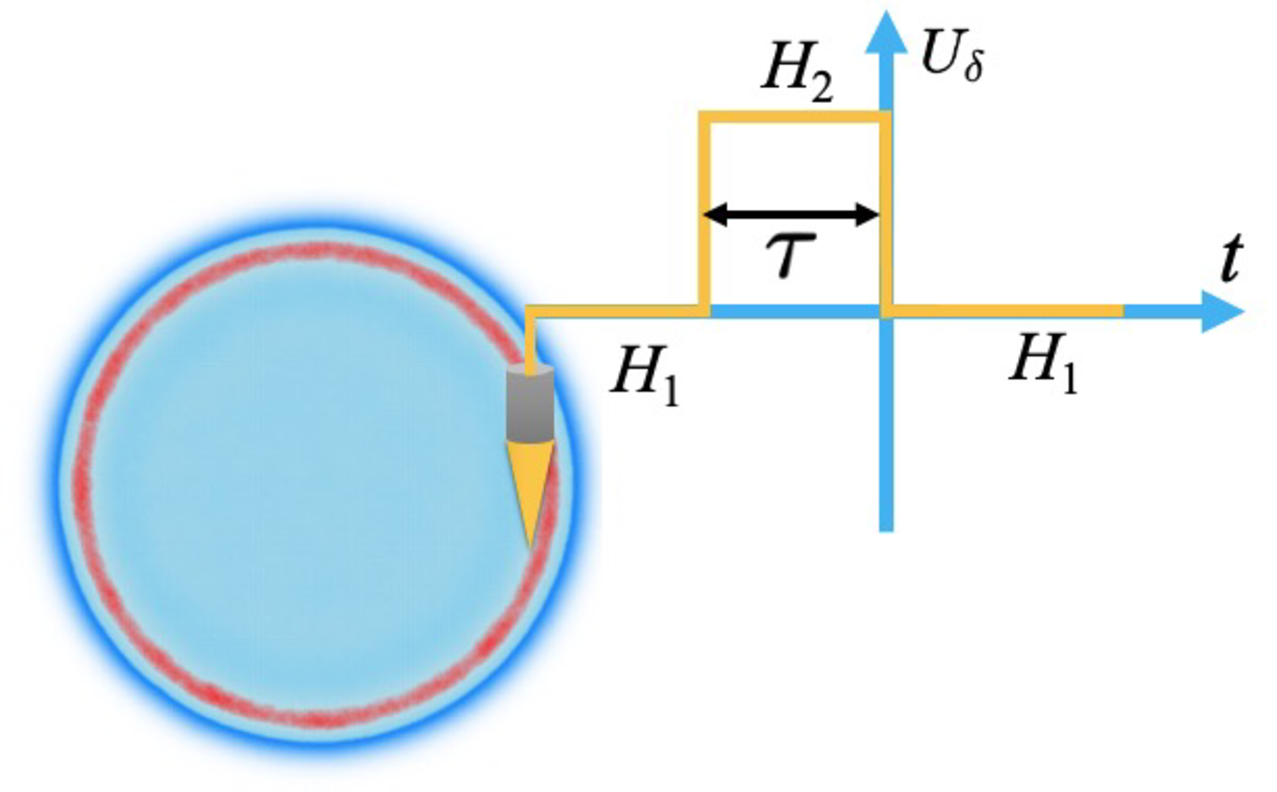}
  \caption{\label{H1H2H1} A potential is applied at the edge of a FQH liquid with a filling factor of $\frac{1}{3}$ to simulate a voltage pulse. For a disk system with $N_e$ electrons, the radius $R$ is given by $\sqrt{2N_{\text{orb}}}$, and the number of orbitals $N_{\text{orb}}$ is equal to $\frac{N_e}{\nu}$. The evolution process can be divided into three parts: Initially, the system is in the ground state of Hamiltonian $H_1$. Subsequently, a potential $V(z)$ is applied at the edge of the system for a duration $\tau$ (unit: $\hbar \varepsilon l_B/e^2$) to simulate the square voltage pulse described in Ref.~\onlinecite{Yusa2022}. During this period, the system's Hamiltonian is $H_2 = H_1 + V(z)$. Finally, the moment when the potential $V(z)$ is removed is denoted as $t=0$, and the system's Hamiltonian reverts to $H_1$.}
\end{figure}
In this work, we consider a two-dimensional electron gas (2DES) system in disk geometry that features an open boundary with a perpendicular magnetic field. The conservation of angular momentum arises from the rotational symmetry, leading to the following single-particle wave function at the lowest Landau level (LLL):
\begin{equation}
  |m\rangle = \frac{1}{\sqrt{2\pi}l_B} \frac{1}{\sqrt{2^m m!}} z^m e^{-|z|^2/4l_B^2}.
\end{equation}
where $z = x + i y$ are the coordinates of the electron in the plane, $l_B = \sqrt{\hbar c/eB}$ is the magnetic length, and $m$ is the angular momentum along the $z$-axis labels the degenerate orbital states in each Landau level. For a finite system with orbitals $N_{orb}$, $m$ takes a value from zero to $N_{orb} - 1$.

As depicted in Fig.~\ref{H1H2H1}, a $\delta$ potential with strength $U_\delta$ at the position $w$ near the boundary mimics the tip potential used in the experiment.
\begin{equation}\label{deltapotential}
    V(z) = U_\delta \delta(z - w)
\end{equation}
While $w$ resides in the bulk, this potential might generate a quasihole excitation~\cite{Hu08} with a fractional charge. In the Hilbert space defined by orbitals $\{ |m\rangle \}$, the LLL projected matrix element $V_{mn} = \langle m | V(z) | n \rangle$ can be expressed as a one-dimensional integral
\begin{eqnarray}
  V_{mn} &=& \frac{U_\delta}{2^{(n-m)/2}} \sqrt{\frac{m!}{n!}} \int_0^{\infty} dk e^{-\frac{k^2}{2}} k^{n-m+1}
  L_{m}^{n-m} \left (\frac{k^2}{2} \right ) \nonumber \\
  &&\times J_{n-m}(k \vert w \vert) \left ( \frac{w}{\vert w \vert} \right )^{m-n}.
\end{eqnarray}
with $L_m^k(k)$ representing the generalized Laguerre polynomial and $J_m(k)$ denoting the Bessel function. The Hamiltonian of the system is 
\begin{equation}
    H_2 = H_1 + V(z)\Theta(t).
\end{equation}
where $\Theta(t)$ denotes the step function, one for $t \in [0, \tau]$ and $0$ otherwise. The second term mimics the tip pulse potential at location $w$, characterized by a duration $\tau$ and an intensity $U_{\delta}$. The first term $H_1$, represents the Hamiltonian for electron-electron interactions that hosts the FQH state as the ground state at specific filling. Here, for simplicity, we consider the Laughlin state~\cite{Laughlin83}, denoted as $|\Psi_1\rangle$, at $\nu = 1/3$, which is the densest zero-energy eigenstate of the short-range hard-core interaction. In Haldane's pseudopotential formalism~\cite{HaldanePP}, this interaction can be described by $V_m=\delta_{1,m}$.  Beginning with $\Psi_1$, applying the tip potential over the time interval $\tau$ leads to the state evolving into
\begin{equation} \label{psi0}
    |\Psi(0)\rangle = \int_0^{\tau} dt \exp(-iH_2 t/\hbar)|\Psi_1\rangle.
\end{equation}
We label this as the initial state at $t = 0$. For $t>0$, the potential $V(z)$ is turned off, causing the system's Hamiltonian to return to $H_1$, and the wave function evolves as
\begin{equation}
  |\Psi(t)\rangle = \int_0^{t} dt‘ \exp(-iH_1t'/\hbar)|\Psi(0)\rangle.
\end{equation}
In the subsequent analysis, we thoroughly investigate the properties of $|\Psi(t)\rangle$.

\section{Quench dynamics at the FQH edge} \label{sec:quench}
\begin{figure}
    \centering
    \includegraphics[width=8cm]{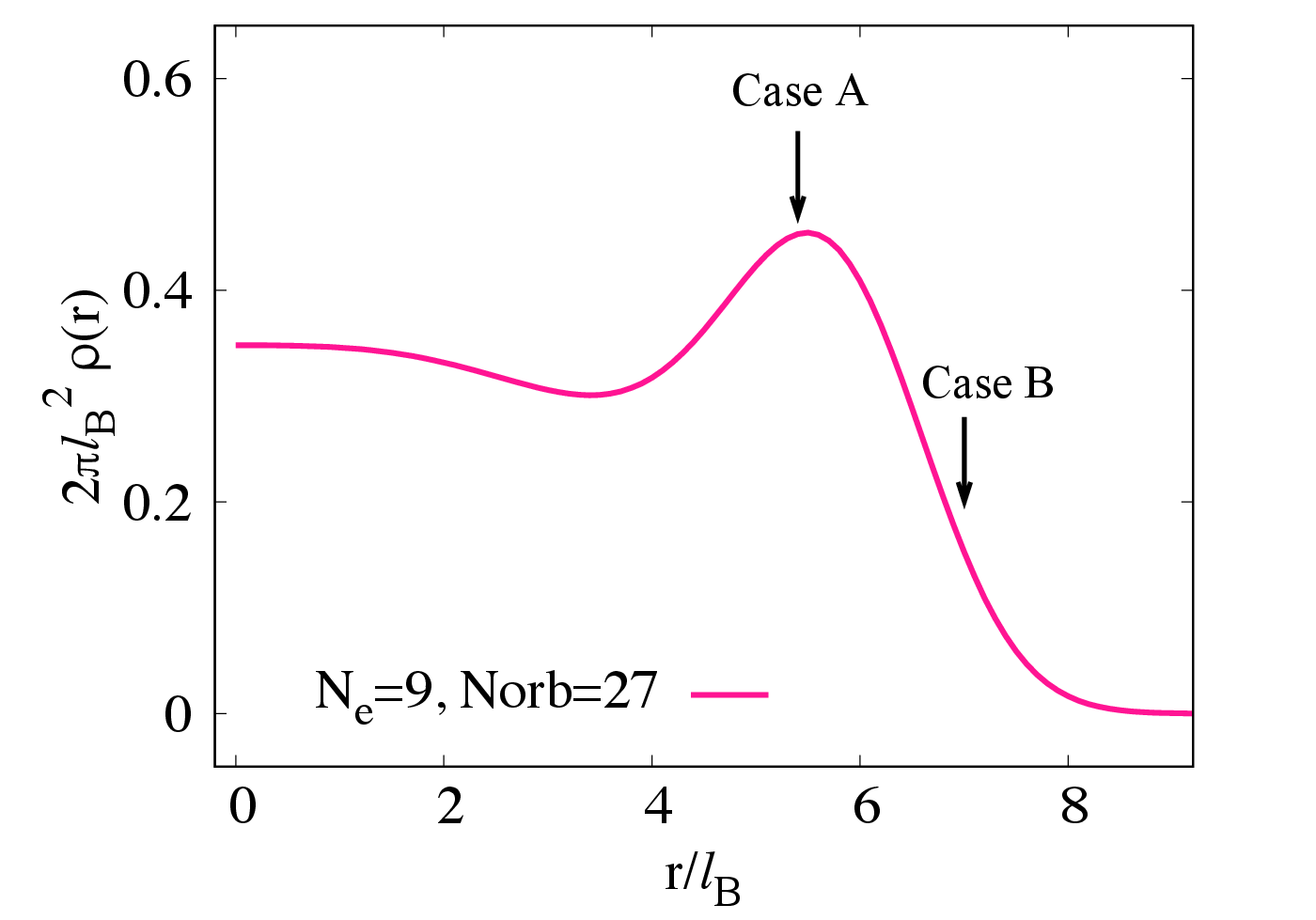}
    \caption{The radial density profile of the Laughlin state for 9 electrons.In two distinct cases, pulse excitations are applied with parameters set to $w=5.4l_B$(Case A) and $w=7.0l_B$(Case B). one corresponds to the spatial location where the edge electron density of the system attains its extremum, while the other is situated in the vicinity of the system boundary.}
    \label{fig:densityN9}
\end{figure}

The presence of boundaries in the system results in a nonuniform edge electron density distribution of the FQH liquid. This non-uniformity originates from the electron-electron correlation and results in an edge dipole momentum, which is related to the Hall viscosity and topological properties of the FQH liquids~\cite{Yejeprb,Yangyiprb23}. Moreover, the edge excitation velocity is defined by the slope of the EMP dispersion in the long-wavelength limit $k \rightarrow 0$~\cite{Huprb09}. For model Hamiltonian with $V_1$ interaction, the edge states are also zero energy eigenstates, and thus the edge velocity is zero. In this scenario, the impact of the tip potential does not propagate along the boundary, meaning that the system does not experience rotation except diffusion over time. To effectively capture the small changes in density evolution, we investigate the time evolution of the electron density diffusion induced by the pulse tip potential with a residual density distribution defined as
\begin{equation}
  \rho(t)=\left \langle \Psi(t) | \Psi(t)  \right \rangle -\rho_1,
\end{equation}
where $\rho_1 = |\Psi_1|^2$ represents the non-perturbed electron density of the Laughlin state prior to the introduction of the pulse tip potential. Due to the tip potential breaks the rotational symmetry, numerical diagonalization can only be applied to relatively small system sizes. For the Laughlin state at $\nu = 1/3$, we consider a system with $N_e = 9$ electrons in $N_{\text{orb}} = 27$ orbitals. Then the disk has a radius around $R \sim \sqrt{2N_{orb}} = 7.3 l_B$. As depicted in Fig.~\ref{fig:densityN9}, the radial density maintains a constant value at the disk's center but becomes inhomogeneous toward the edge, reaching a peak around $r \sim 5.4l_B$.

\begin{figure}
  \includegraphics[width=8cm]{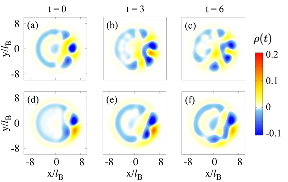}
  \caption{\label{density}Evolution of the residual density under quenched dynamics in a system comprising $9$ electrons in $27$ orbitals, with $\tau = 3$ and $U_{\delta} = 1$. (a–c) depict the dynamic process of edge density diffusing into the bulk over time when the tip potential is at $w = 5.4l_B$; (d–f) correspond to $w = 7.0l_B$, where the penetration depth of edge density into the bulk is significantly greater than that at $w = 7.0l_B$ on the same time scale, demonstrating that the pulse position modulates the diffusion intensity.
  }
\end{figure}

The pulse tip potential is established with a fixed strength of $U_{\delta} = 1$ and a duration $\tau = 3$. Two values are assigned to position $w$, namely $w = 5.4l_B$ (case A) and $w = 7.0l_B$ (case B). One corresponds to the density maximum, while the other is situated near the boundary. As shown in Fig.~\ref{density}, we examine $\rho(t)$ for both cases at several time points. In case A, illustrated in (a)-(c), when the tip is positioned at the density peak, the residual electron density clearly shows diffusion traits into the bulk. In contrast, as the tip nears the edge, as in case B shown in (d)-(f), $\rho(t)$ appears to be more concentrated along the edge. This aligns with the experimental findings described in Ref.~\onlinecite{Yusa2022}, indicating that the pulse near the edge has the capability to excite both edge waves and bulk modes. However, our simulation reveals that the influence of these two excitations highly depends on the tip position $w$, as analyzed below.

\begin{figure}[ht]
  \includegraphics[width=8.5cm]{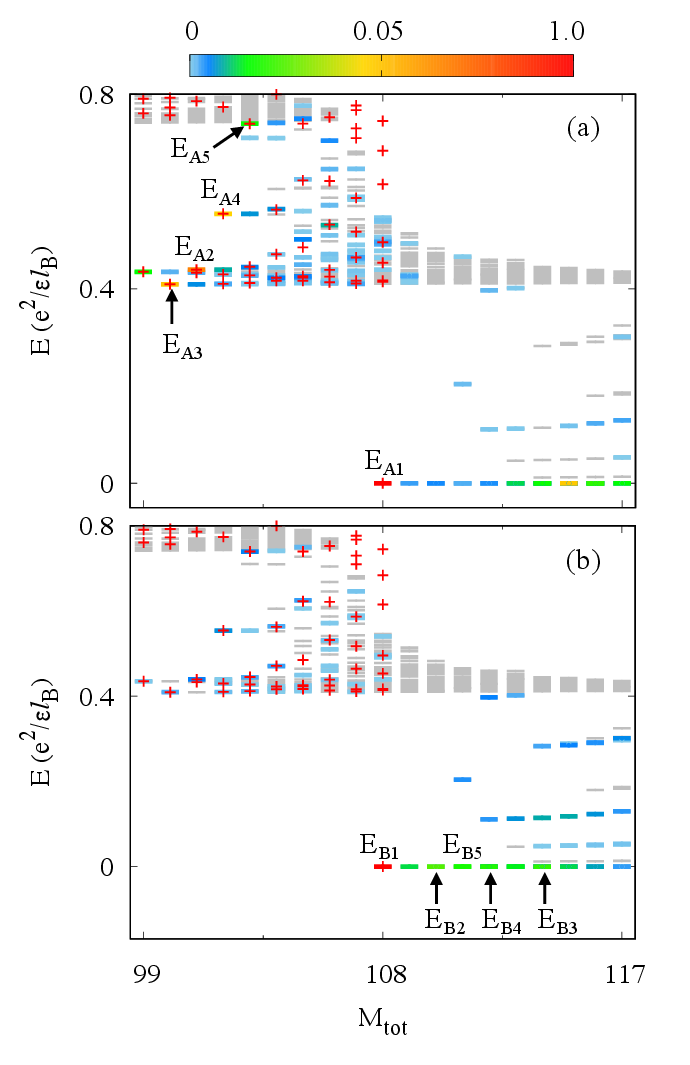}
  \caption{\label{Eoverlap}Energy spectrum for 9 electrons and 27 orbitals. Color bar indicate the overlap of edge states, bulk states, and $\Psi(0)$. The red cross marks      in the same system. (a) Case A, with the states having higher weights predominantly distributed in the bulk states. (b) Case B, with the states having higher weights predominantly distributed in the edge states.}
\end{figure}

Next, we conduct a further investigation of this dynamical process from the perspective of wave functions. Within the framework of this study, the initial state for system evolution is the Laughlin state $\Psi_1$. Taking a system with the electron number $N_e = 9$ as an example, $\Psi_1$ corresponds to a zero-energy eigenstate with angular momentum $M_{0} = 3N_e(N_e - 1)/2 = 108$. The pulse process, accompanied by energy injection into the system, may induce transitions to high-energy excited states. To quantitatively characterize the quantum transition process from ground state to excited states after pulse application, Fig.~\ref{Eoverlap} presents the energy spectrum of the $V_1$ Hamiltonian for a system with $N_{orb} = 27$ orbitals. A color bar illustrates the squared overlap distributions between the post-pulse state $\Psi(0)$ for case A (Fig.(a)) or B (Fig.(b)) and every state in the energy spectrum. It is shown that the nonzero overlap states are distributed on both sides of the ground state. The numbers indicated in the figure correspond to the square overlap values, which are the largest five states. The states indicated by the red cross points are the eigenstates in the subspace of zero center of mass (COM) angular momentum. The COM operator is characterized as $\hat{M} = B^\dagger B$, where the ladder operator is given by $\hat{B} = \sum_i b_i/N_e$, with $b_i$ representing the ladder operator for each guiding center orbital. Our earlier research~\cite{Wu-Qing} demonstrates that the neutral magnetoroton excitation mode lives in this subspace, particularly focusing on low-energy states with angular momentum $M_{tot}$ in the range $[M_0 - N_e, M_0]$. In case A, as illustrated in Fig.~\ref{Eoverlap}(a), all states with the highest overlap, except the ground state, are located in the left segment of the spectrum among the bulk excited states. In particular, these states with maximal overlap are located within the $M = 0$ subspace and are marked with red cross symbols. They predominantly occupy the low-energy part of the spectrum on this side, implying a significant contribution from the magnetoroton excitation mode. In addition, certain states demonstrating the highest overlap are located in the upper segment of the energy spectrum, such as the $E_{A4}$ state. This excitation could be explained by magnetoroton excitation, where a particle transitions to a higher $\Lambda$ level within the framework of composite fermion theory~\cite{YY25PRB}. In case B, when the tip is located near the boundary, as shown in Fig.~\ref{Eoverlap}(b), the states exhibiting the highest overlap are found among the zero energy states in the right part of the spectrum, particularly associated with the edge excitation mode of the FQH liquid~\cite{Wen90}. The variation observed in the weight distribution illustrates that the excitation of the pulse is significantly influenced by the tip's position.

\begin{figure}
  \includegraphics[width=8.5cm]{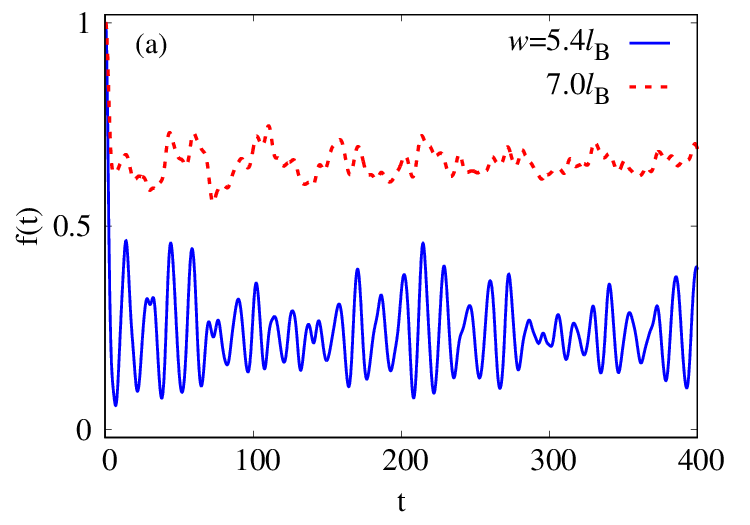}
  \includegraphics[width=8.5cm]{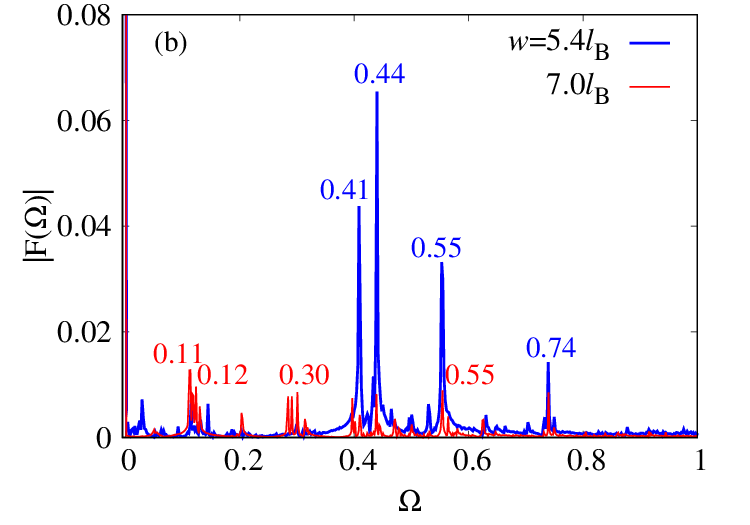}
  \caption{\label{Toverlap}In a 9-electron 27-orbital system with $w=5.4l_B$, $\tau=3$, and $U_{\delta}=1$, the evolution of fidelity $f(t)$ is presented. The inset illustrates the residual electron density distributions at $t=0$, $7$, and $15$, demonstrating a periodic evolution pattern: the density wavepacket diffuses from the initial edge region $t = 0$ into the bulk $t = 7$ , followed by re-emerging edge accumulation upon completion of the first oscillation cycle ($t=15$).(b) The Fourier frequency spectrum of the fidelity in Fig.~\ref{Toverlap}. The inset shows the parts with larger projection weights in Fig.~\ref{Eoverlap}, labeled in descending order of weight as $E_1, E_2, ...$, and these states are also marked by the magnetoroton energy spectrum.}
\end{figure}

Now we calculate the post-quench fidelity $f(t)=|\langle \Psi(0)|\Psi(t)\rangle|^2$ to monitor the dynamical evolution. The oscillation period of $f(t)$ is determined by the energy difference between the eigenstates in the energy spectrum. The oscillation amplitude is influenced by the overlap between the initial state $\Psi(0)$ and the eigenstates of the Hamiltonian $H_1$. Analytically, the fidelity $f(t)$ can be expanded in terms of energy eigenstates as:  
\begin{align}
&|\langle \Psi(0) | \Psi(t) \rangle|^2 
= \left| \sum_i |c_i|^2 e^{-i E_i t/\hbar} \right|^2 \nonumber \\
&= \sum_i |c_i|^4 + \sum_{i < j} 2 |c_i|^2 |c_j|^2 \cos\left( \frac{(E_i - E_j)t}{\hbar} \right).
\end{align}

As shown in Fig.~\ref{Toverlap}(a), $f(t)$ exhibits distinct multiperiod oscillation characteristics. During the initial stage, the fidelity decreases monotonically as the electron density perturbation diffuses into the bulk or along the edge from the tip. After a certain time, the fidelity begins to oscillate, indicating that the electron density wavepacket has re-emerged at the edge. This oscillation behavior is consistent with the periodic diffusion of the electron density wavepacket observed in Fig.~\ref{density}. The oscillation period is determined by the energy differences between the eigenstates involved in the dynamics. In case A, where the tip is positioned at $w = 5.4l_B$, the oscillation period is approximately $T \approx 14.3$ (in units of $\hbar \varepsilon l_B/e^2$) corresponding to a frequency of $\Omega \approx 0.44$ which is very close to the energy of the magnetoroton mode in large $k$-limit. The amplitude of these oscillations is relatively large, indicating that the excitation energy is high and that the system diffuses during this process. In contrast, for case B, the fidelity exhibits larger values and slower oscillations, suggesting a greater retention of ground state characteristics, consistent with the tip's proximity to the edge. Furthermore, the observation of long-period oscillations in $f(t)$ indicates lower energy excitations at the FQH edge.

To extract the characteristic oscillation frequencies, we performed a discrete Fourier transform of the fidelity $f(t)$ from Fig.~\ref{Toverlap}(a), with the resulting frequency spectrum $F(\Omega)$ shown in Fig.~\ref{Toverlap}(b). The inset identifies the five dominant energy levels ($E_{A1}$ to $E_{A5}$) from Fig.~\ref{Eoverlap} with the highest integral overlap weights, ranked in descending order. The spectral peaks in $F(\Omega)$ exhibit a correspondence with the energy differences between these levels, as detailed in Table~\ref{deltaE}. For example, the peak of the fundamental frequency at $\hbar\Omega_{12} = |E_{A1} - E_{A2}| = 0.44$ corresponds to a period $T_{12} = 2\pi/\Omega_{12} \approx 14.3$, and the secondary peak at $\hbar\Omega_{13} = |E_{A1} - E_{A3}| = 0.41$ yields $T_{13} = 2\pi/\Omega_{13} \approx 15.3$. These periodicities quantitatively explain the dominant oscillations observed in Fig.~\ref{Toverlap}. Moreover, all identified dominant energy levels coincide with the magnetoroton spectrum in case A, confirming that the oscillation periods of $f(t)$ are governed by these transitions. This indicates that after the edge pulse excitation in the Hall fluid, the electron diffusion into the bulk is directly modulated by the collective modes of the magnetoroton excitations.

\begin{table}
  \begin{center}
    \caption{\label{deltaE}The table of absolute value of the energy difference $|E_i - E_j|$ marked in Fig.~\ref{Eoverlap}, where the underlined energy differences correspond to the peak frequencies in the Fourier spectrum.}
    \begin{tabular}{c c c c c c c}
      \hline \hline
                          & $E_{A1}$   \hspace{0.3cm}            & $E_{A2}$ \hspace{0.3cm} & $E_{A3}$ \hspace{0.3cm} & $E_{A4}$ \hspace{0.3cm} & $E_{A5}$ \hspace{0.3cm} \\
      \hline
      $E_{A1}$\hspace{0.3cm} & 0    \hspace{0.3cm}              &        &        &        &       \\
      $E_{A2}$\hspace{0.3cm} & \underline{0.44} \hspace{0.3cm} & 0    \hspace{0.3cm}  &        &        &       \\
      $E_{A3}$\hspace{0.3cm} & \underline{0.41} \hspace{0.3cm} & 0.03 \hspace{0.3cm} & 0   \hspace{0.3cm}   &        &       \\
      $E_{A4}$\hspace{0.3cm} & \underline{0.55} \hspace{0.3cm} & 0.11 \hspace{0.3cm} & 0.14 \hspace{0.3cm} & 0  \hspace{0.3cm}    &       \\
      $E_{A5}$\hspace{0.3cm} & \underline{0.74} \hspace{0.3cm} & 0.30 \hspace{0.3cm} & 0.33 \hspace{0.3cm} & 0.19 \hspace{0.3cm} & 0 \hspace{0.3cm}    \\
      \hline \hline
    \end{tabular}
  \end{center}
\end{table}
\textit{Effect of Coulomb interaction}
In the above analysis, we have considered the short-range interaction $V_1$ in the Hamiltonian, which is a good approximation for the Laughlin state. However, in reality, the Coulomb interaction leads to a more complex energy spectrum and can modify the excitation energies of the system. First of all, the edge states are no longer zero energy eigenstates, and thus the edge velocity is non-zero. Therefore the density modulations induced by the tip potential can propagate along the edge, leading to a more complex quench dynamics. Moreover, the Coulomb interaction also leads to a more complex bulk energy spectrum, which can affect the oscillation period of the fidelity. To investigate this effect, we have performed a numerical diagonalization of the Hamiltonian with Coulomb interaction for the same system size as above. The results show that the energy spectrum becomes more complex, and the oscillation period of the fidelity is slightly modified. However, the overall qualitative behavior remains similar to that observed with the short-range interaction. This indicates that the short-range interaction is a good approximation for studying the quench dynamics at the edge of FQH liquids.

\section{Analyze of the details of the tip}
\label{sec:factor}

Previously, we discussed the quench dynamics of the edge state induced by a specific pulse potential at the edge of a FQH liquid. The results indicate that the pulse can excite both edge and bulk states, and the excitation characteristics are significantly influenced by the position of the pulse. In this section, we will further analyze the effects of pulse position $w$, duration $\tau$, and intensity $U_{\delta}$ on excitation characteristics. As shown above, the pulse position is particularly crucial as it determines the initial electron density distribution, which in turn affects the excitation process. The duration and intensity of the pulse also play an important role in determining the amplitude and energy of the excitation. By systematically varying these parameters, we can gain a deeper understanding of how they influence the quench dynamics and the resulting excitation characteristics. 

\begin{figure}
  \includegraphics[width=8cm]{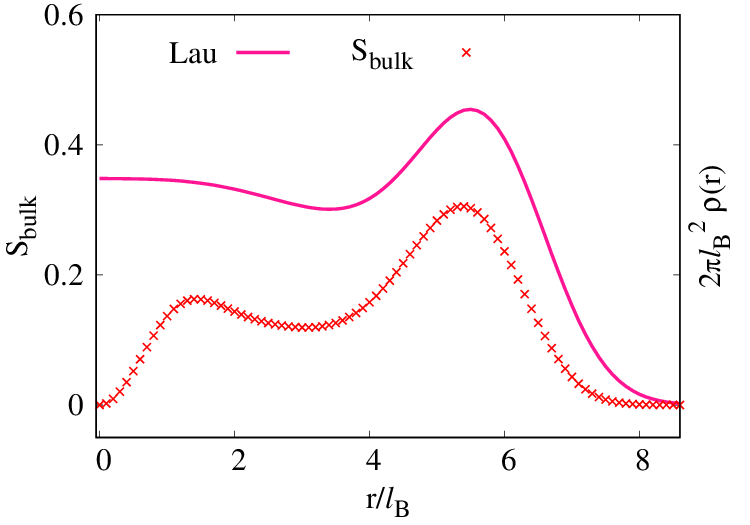}
  \caption{\label{Sbulk}The dependence of $S_{\text{bulk}}$ on the pulse position. The peak location of electron density nearly coincides with the maximal value of $S_{\text{bulk}}$. This correspondence indicates that positioning the excitation pulse near the electron density peak effectively enhances the hybridization effect in the bulk state.
  }
\end{figure}

\textit{Pulse positon.}
To quantitatively analyze the contribution ratio of these two types of state to the time evolution, we introduce the bulk contribution $S_{\text{bulk}}$ as follows: 
\begin{equation}
S_{\text{bulk}} = \sum_{m=M_0-N_e}^{M_0-1} \sum_{i=0}^{39} \left| \langle \Psi(0) | \phi_{m,i} \rangle \right|^2
\end{equation}
where $\phi_{m,i}$ represents the $i$-th excited state at angular momentum $m$, with double summation traversing the lowest $40$ excited states in each angular momentum sector within the $[M_0-N_e, M_0-1]$ angular momentum window where the magnetoroton excitation lives.The larger the value of $S_{\text{bulk}}$, the more readily the initial state $\Psi(0)$ can be excited into the bulk states within this angular momentum interval, indicating that the system evolution will be significantly governed by the collective modes of magnetoroton.
Based on the aforementioned analysis, the pulse position may exert a significant regulatory effect on the diffusion dynamics. To investigate this, we systematically varied the spatial position of the excitation pulse, calculated the dependence of $S_{\text{bulk}}$ on the pulse position, and plotted the results in Fig.~\ref{Sbulk}. As shown in the figure, a prominent maximum of $S_{\text{bulk}}$ appears at $r=5.4l_B$, indicating a significant enhancement effect of the pulse position on the diffusion process at this location. 
Furthermore, the distribution of electron density before the pulse, depicted by the solid line in Fig.~\ref{Sbulk}, reveals that the peak region of the electron density near the edge coincides with the maximum of $S_{\text{bulk}}$. When the pulse position moves toward the edge of the system to $r=8.0l_B$, the intensity of $S_{\text{bulk}}$ gradually decreases to zero, which is consistent with the electron density reaching the physical boundary of the system at this position. In particular, the spatial span from the maximum point to the zero point corresponds exactly to the radius of a quasi-hole, that is, $r \simeq 2.5l_B$~\cite{Andrei2014,Liqi_2015,liuzhao,jieli}.

\begin{figure}
  \includegraphics[width=8cm]{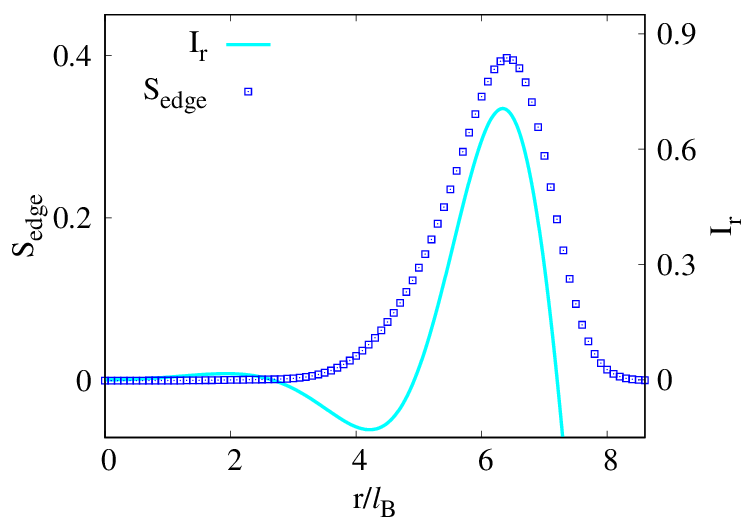}
  \caption{\label{Sedge}The pulse position dependence of $S_{\text{edge}}$. $S_{\text{edge}}$ exhibits an asymmetric unimodal distribution between $3.2l_B$ and $8.2l_B$, peaking at $r=6.4l_B$. The solid line represents the dipole moment $I_{r}$, also maximized at the same location.
  }
\end{figure}

Meanwhile, we can also introduce the edge contribution $S_{\text{edge}}$ to quantify the contribution of the edge state to the quench dynamics. Edge states are defined as zero-energy eigenstates in the right part of the energy spectrum, which are typically located in the angular momentum range $[M_0 + 1, M_0 + N_e]$. The edge contribution is defined as
\begin{equation}
S_{\text{edge}} = \sum_{m=M_0+1}^{M_0+N_e} \sum_{i=0}^{n} \left| \langle \Psi(0) | \phi_{m,i} \rangle \right|^2
\end{equation}
The truncation parameter $n$ is determined by the energy gap between the ground state and the bulk states. For the $V_1 = 1$ model Hamiltonian, the energy gap $\Delta E \approx 0.41$, and $n$ is taken as the number of edge states with energies below $\Delta E$. The parameter $S_{\text{edge}}$ quantifies the percentage proportion of edge states involved in the quench dynamics after the termination of the pulse. The ratio between $S_{\text{bulk}}$ and $S_{\text{edge}}$ directly reflects the competitive relationship between the bulk and edge states during the evolution process.
Similarly, we calculated the dependence of $S_{\text{edge}}$ on the pulse position, as shown in Fig.~\ref{Sedge}. Unlike $S_{\text{bulk}}$, $S_{\text{edge}}$ exhibits accumulation only within a specific edge-confined region. Specifically, as the probe moves from $r = 3.2l_B$ toward the edge, $S_{\text{edge}}$ increases monotonically, reaches a maximum at $r = 6.4l_B$, then rapidly decays to zero at $r = 8.2l_B$.  This indicates that the pulse position significantly influences the excitation of edge states, with the maximum excitation occurring at $r = 6.4l_B$.  In order to further understand the position of the maxima excitation, we analyze the dipole moment of the edge state. The dipole moment is a measure of the asymmetry of the electron density distribution at the edge, which is closely related to the quasiparticle excitation. It is defined as $I_{r} = \int _0^{r}r(\rho(r)-\nu)dr$. As shown in Fig.~\ref{Sedge}, it is interesting that the dipole moment $I_{r}$ also exhibits a maximum at $r = 6.4l_B$, which coincides with the position of the maximum contribution of the edge excitations. This indicates that the pulse position at $r = 6.4l_B$ corresponds to the point where the electron density distribution is most asymmetric, leading to particle-hole pairing excitation near the edge. Additionally, the $S_{\text{edge}}$ extends roughly $5l_B$, matching the quasihole diameter or the particle-hole pair's central distance~\cite{Andrei2014,Liqi_2015,liuzhao,jieli}.

In conclusion, the pulse position dependence of $S_{\text{edge}}$ indicates that the excitation of edge states is highly sensitive to the position of the pulse. When the pulse is applied at the edge, the edge states are excited, which is closely related to the quasiparticle excitation, leading to a significant increase in $S_{\text{edge}}$. In contrast, when the pulse is applied to the center of the disk, the edge states are not excited and $S_{\text{edge}}$ is zero.

\begin{figure}
  \includegraphics[width=8cm]{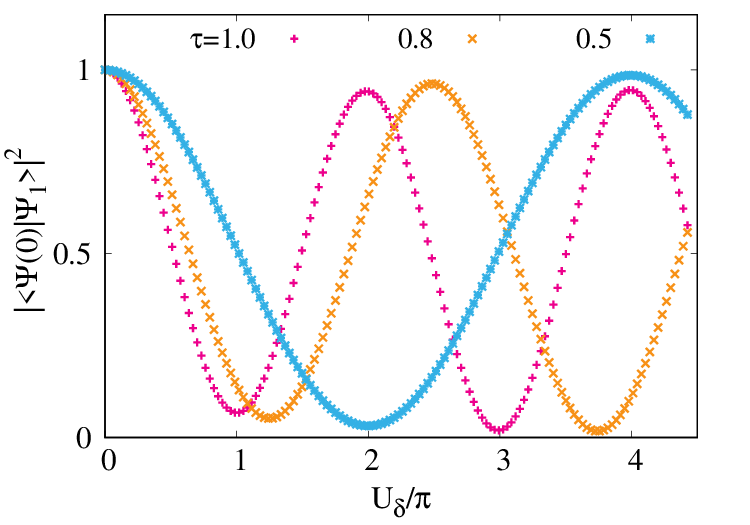}
  \caption{\label{Lau_UTau} Variations of the wavefunction overlap $|\left \langle\Psi(0)|\Psi_1\right \rangle|^2$ as a function of the pulse strength $U_{\delta}$ for various $\tau$. }
\end{figure}

\textit{Pulse Duration and strength.}
In the experiment, pulse duration $\tau$ and strength $U_{\delta}$ are adjustable parameters. We analyzed the effect of pulse duration on excitation by calculating $|\langle\Psi(0)|\Psi_1\rangle|^2$ versus $U_{\delta}$ for various $\tau$. Fig.~\ref{Lau_UTau} shows that $|\langle\Psi(0)|\Psi_1\rangle|^2$ varies periodically with $U_{\delta}$ with a slowly decaying amplitude, particularly for small $\tau$. Moreover, $\tau$ and the oscillation period $\Delta U_{\delta}$ show an inverse relationship: $\Delta U_{\delta} \cdot \tau = 2\pi$. Treating all excited states, excluding only the ground state $|\Psi_1\rangle$ before applying the pulse, as a single system excitation, this oscillation is very likely the Rabi oscillation in a two-level system with a gap $U_{\delta}$.
The transition probability is proportional to $|\sin(U_{\delta} \tau /2)|^2$ and thus $U_{\delta}$ is periodic with $2\pi\hbar/\tau$ for a fixed $\tau$. Thus, the ground state excitation can be adjusted by the pulse duration $\tau$ and the strength $U_{\delta}$.

\section{Discussions and Conclusions}
\label{sec:dis}
In this work, we have investigated the quench dynamics of edge excitations in FQH liquid at the filling factor $\nu = 1/3$ motivated by recent pump probe reflectance measurements in time-resolved photoluminescence spectroscopy. By applying a tip potential to the edge of the Hall liquid, we simulate an electrical pulse that excites the Laughlin state into the energy spectrum branch of the edge mode and magnetic rotor. The electron density evolution reveals that due to chiral edge modes, electrons move along the edge while also diffusing into the bulk. We calculated the fidelity after quenching and performed spectral analysis on the time-dependent fidelity. The results show that the evolution frequency aligns with the energy gap of the magnetic rotor, and electron diffusion into the bulk is attributed to the winding of the magnetic rotor. In addition, we discussed how the location, intensity, and duration of the pulse affect the quench dynamics. The results indicate that the position of the pulse significantly influences the excitation characteristics, with the maximum excitation occurring when the pulse is applied at the peak of the electron density. The duration and intensity of the pulse also play crucial roles in determining the amplitude and energy of the excitation. The periodic nature of the excitation suggests that the system can be tuned to achieve optimal excitation conditions by adjusting these parameters. In conclusion, our study provides a comprehensive understanding of the quench dynamics of edge excitations in FQH liquids. The results highlight the importance of the pulse position, duration, and intensity in controlling the excitation characteristics. This work initiates further investigations into the dynamics of edge excitations in more exotic FQH systems and their potential applications in quantum information processing and quantum simulation.

\acknowledgments
We thank Han-Tao Lu for the help of the time-dependent Lanczos algorithm. This work was supported by the National Natural Science Foundation of China Grants No. 12474140 and No. 12347101. C-X. Jiang acknowledges the support of the China Scholarship Council Grant No. 202406050101. 

\bibliography{biblio_edge.bib}

\end{document}